\RequirePackage{lineno}  
\documentclass[aps,prl,twocolumn,showpacs,superscriptaddress,groupedaddress]{revtex4}  
\usepackage{graphicx}  
\usepackage{dcolumn}   
\usepackage{bm}        
\usepackage{amssymb}   
\usepackage{color}

\newcommand{\met}{\ensuremath{E\!\!\!/_T}}
\newcommand{\iso}{\ensuremath{I}}
\newcommand{\dstar}{\ensuremath{D^{*}(2010)^+}}
\newcommand{\ds}{\ensuremath{D^{*+}}}

\begin{document}



\title{ Measurement of vector boson plus $\dstar$ meson production 
in $\bar{p}p$ collisions at $\sqrt{s}=1.96\, {\rm TeV}$}
\affiliation{Institute of Physics, Academia Sinica, Taipei, Taiwan 11529, Republic of China}
\affiliation{Argonne National Laboratory, Argonne, Illinois 60439, USA}
\affiliation{University of Athens, 157 71 Athens, Greece}
\affiliation{Institut de Fisica d'Altes Energies, ICREA, Universitat Autonoma de Barcelona, E-08193, Bellaterra (Barcelona), Spain}
\affiliation{Baylor University, Waco, Texas 76798, USA}
\affiliation{Istituto Nazionale di Fisica Nucleare Bologna, \ensuremath{^{kk}}University of Bologna, I-40127 Bologna, Italy}
\affiliation{University of California, Davis, Davis, California 95616, USA}
\affiliation{University of California, Los Angeles, Los Angeles, California 90024, USA}
\affiliation{Instituto de Fisica de Cantabria, CSIC-University of Cantabria, 39005 Santander, Spain}
\affiliation{Carnegie Mellon University, Pittsburgh, Pennsylvania 15213, USA}
\affiliation{Enrico Fermi Institute, University of Chicago, Chicago, Illinois 60637, USA}
\affiliation{Comenius University, 842 48 Bratislava, Slovakia; Institute of Experimental Physics, 040 01 Kosice, Slovakia}
\affiliation{Joint Institute for Nuclear Research, RU-141980 Dubna, Russia}
\affiliation{Duke University, Durham, North Carolina 27708, USA}
\affiliation{Fermi National Accelerator Laboratory, Batavia, Illinois 60510, USA}
\affiliation{University of Florida, Gainesville, Florida 32611, USA}
\affiliation{Laboratori Nazionali di Frascati, Istituto Nazionale di Fisica Nucleare, I-00044 Frascati, Italy}
\affiliation{University of Geneva, CH-1211 Geneva 4, Switzerland}
\affiliation{Glasgow University, Glasgow G12 8QQ, United Kingdom}
\affiliation{Harvard University, Cambridge, Massachusetts 02138, USA}
\affiliation{Division of High Energy Physics, Department of Physics, University of Helsinki, FIN-00014, Helsinki, Finland; Helsinki Institute of Physics, FIN-00014, Helsinki, Finland}
\affiliation{University of Illinois, Urbana, Illinois 61801, USA}
\affiliation{The Johns Hopkins University, Baltimore, Maryland 21218, USA}
\affiliation{Institut f\"{u}r Experimentelle Kernphysik, Karlsruhe Institute of Technology, D-76131 Karlsruhe, Germany}
\affiliation{Center for High Energy Physics: Kyungpook National University, Daegu 702-701, Korea; Seoul National University, Seoul 151-742, Korea; Sungkyunkwan University, Suwon 440-746, Korea; Korea Institute of Science and Technology Information, Daejeon 305-806, Korea; Chonnam National University, Gwangju 500-757, Korea; Chonbuk National University, Jeonju 561-756, Korea; Ewha Womans University, Seoul, 120-750, Korea}
\affiliation{Ernest Orlando Lawrence Berkeley National Laboratory, Berkeley, California 94720, USA}
\affiliation{University of Liverpool, Liverpool L69 7ZE, United Kingdom}
\affiliation{University College London, London WC1E 6BT, United Kingdom}
\affiliation{Centro de Investigaciones Energeticas Medioambientales y Tecnologicas, E-28040 Madrid, Spain}
\affiliation{Massachusetts Institute of Technology, Cambridge, Massachusetts 02139, USA}
\affiliation{University of Michigan, Ann Arbor, Michigan 48109, USA}
\affiliation{Michigan State University, East Lansing, Michigan 48824, USA}
\affiliation{Institution for Theoretical and Experimental Physics, ITEP, Moscow 117259, Russia}
\affiliation{University of New Mexico, Albuquerque, New Mexico 87131, USA}
\affiliation{The Ohio State University, Columbus, Ohio 43210, USA}
\affiliation{Okayama University, Okayama 700-8530, Japan}
\affiliation{Osaka City University, Osaka 558-8585, Japan}
\affiliation{University of Oxford, Oxford OX1 3RH, United Kingdom}
\affiliation{Istituto Nazionale di Fisica Nucleare, Sezione di Padova, \ensuremath{^{ll}}University of Padova, I-35131 Padova, Italy}
\affiliation{University of Pennsylvania, Philadelphia, Pennsylvania 19104, USA}
\affiliation{Istituto Nazionale di Fisica Nucleare Pisa, \ensuremath{^{mm}}University of Pisa, \ensuremath{^{nn}}University of Siena, \ensuremath{^{oo}}Scuola Normale Superiore, I-56127 Pisa, Italy, \ensuremath{^{pp}}INFN Pavia, I-27100 Pavia, Italy, \ensuremath{^{qq}}University of Pavia, I-27100 Pavia, Italy}
\affiliation{University of Pittsburgh, Pittsburgh, Pennsylvania 15260, USA}
\affiliation{Purdue University, West Lafayette, Indiana 47907, USA}
\affiliation{University of Rochester, Rochester, New York 14627, USA}
\affiliation{The Rockefeller University, New York, New York 10065, USA}
\affiliation{Istituto Nazionale di Fisica Nucleare, Sezione di Roma 1, \ensuremath{^{rr}}Sapienza Universit\`{a} di Roma, I-00185 Roma, Italy}
\affiliation{Mitchell Institute for Fundamental Physics and Astronomy, Texas A\&M University, College Station, Texas 77843, USA}
\affiliation{Istituto Nazionale di Fisica Nucleare Trieste, \ensuremath{^{ss}}Gruppo Collegato di Udine, \ensuremath{^{tt}}University of Udine, I-33100 Udine, Italy, \ensuremath{^{uu}}University of Trieste, I-34127 Trieste, Italy}
\affiliation{University of Tsukuba, Tsukuba, Ibaraki 305, Japan}
\affiliation{Tufts University, Medford, Massachusetts 02155, USA}
\affiliation{University of Virginia, Charlottesville, Virginia 22906, USA}
\affiliation{Waseda University, Tokyo 169, Japan}
\affiliation{Wayne State University, Detroit, Michigan 48201, USA}
\affiliation{University of Wisconsin, Madison, Wisconsin 53706, USA}
\affiliation{Yale University, New Haven, Connecticut 06520, USA}

\author{T.~Aaltonen}
\affiliation{Division of High Energy Physics, Department of Physics, University of Helsinki, FIN-00014, Helsinki, Finland; Helsinki Institute of Physics, FIN-00014, Helsinki, Finland}
\author{S.~Amerio\ensuremath{^{ll}}}
\affiliation{Istituto Nazionale di Fisica Nucleare, Sezione di Padova, \ensuremath{^{ll}}University of Padova, I-35131 Padova, Italy}
\author{D.~Amidei}
\affiliation{University of Michigan, Ann Arbor, Michigan 48109, USA}
\author{A.~Anastassov\ensuremath{^{w}}}
\affiliation{Fermi National Accelerator Laboratory, Batavia, Illinois 60510, USA}
\author{A.~Annovi}
\affiliation{Laboratori Nazionali di Frascati, Istituto Nazionale di Fisica Nucleare, I-00044 Frascati, Italy}
\author{J.~Antos}
\affiliation{Comenius University, 842 48 Bratislava, Slovakia; Institute of Experimental Physics, 040 01 Kosice, Slovakia}
\author{G.~Apollinari}
\affiliation{Fermi National Accelerator Laboratory, Batavia, Illinois 60510, USA}
\author{J.A.~Appel}
\affiliation{Fermi National Accelerator Laboratory, Batavia, Illinois 60510, USA}
\author{T.~Arisawa}
\affiliation{Waseda University, Tokyo 169, Japan}
\author{A.~Artikov}
\affiliation{Joint Institute for Nuclear Research, RU-141980 Dubna, Russia}
% ...

\collaboration{CDF Collaboration}
\altaffiliation[With visitors from]{
\ensuremath{^{a}}University of British Columbia, Vancouver, BC V6T 1Z1, Canada,
\ensuremath{^{b}}Istituto Nazionale di Fisica Nucleare, Sezione di Cagliari, 09042 Monserrato (Cagliari), Italy,
\ensuremath{^{c}}University of California Irvine, Irvine, CA 92697, USA,
\ensuremath{^{d}}Institute of Physics, Academy of Sciences of the Czech Republic, 182~21, Czech Republic,
\ensuremath{^{e}}CERN, CH-1211 Geneva, Switzerland,
\ensuremath{^{f}}Cornell University, Ithaca, NY 14853, USA,
\ensuremath{^{g}}University of Cyprus, Nicosia CY-1678, Cyprus,
\ensuremath{^{h}}Office of Science, U.S. Department of Energy, Washington, DC 20585, USA,
\ensuremath{^{i}}University College Dublin, Dublin 4, Ireland,
\ensuremath{^{j}}ETH, 8092 Z\"{u}rich, Switzerland,
\ensuremath{^{k}}University of Fukui, Fukui City, Fukui Prefecture, Japan 910-0017,
\ensuremath{^{l}}Universidad Iberoamericana, Lomas de Santa Fe, M\'{e}xico, C.P. 01219, Distrito Federal,
\ensuremath{^{m}}University of Iowa, Iowa City, IA 52242, USA,
\ensuremath{^{n}}Kinki University, Higashi-Osaka City, Japan 577-8502,
\ensuremath{^{o}}Kansas State University, Manhattan, KS 66506, USA,
\ensuremath{^{p}}Brookhaven National Laboratory, Upton, NY 11973, USA,
\ensuremath{^{q}}Istituto Nazionale di Fisica Nucleare, Sezione di Lecce, Via Arnesano, I-73100 Lecce, Italy,
\ensuremath{^{r}}Queen Mary, University of London, London, E1 4NS, United Kingdom,
\ensuremath{^{s}}University of Melbourne, Victoria 3010, Australia,
\ensuremath{^{t}}Muons, Inc., Batavia, IL 60510, USA,
\ensuremath{^{u}}Nagasaki Institute of Applied Science, Nagasaki 851-0193, Japan,
\ensuremath{^{v}}National Research Nuclear University, Moscow 115409, Russia,
\ensuremath{^{w}}Northwestern University, Evanston, IL 60208, USA,
\ensuremath{^{x}}University of Notre Dame, Notre Dame, IN 46556, USA,
\ensuremath{^{y}}Universidad de Oviedo, E-33007 Oviedo, Spain,
\ensuremath{^{z}}CNRS-IN2P3, Paris, F-75205 France,
\ensuremath{^{aa}}Universidad Tecnica Federico Santa Maria, 110v Valparaiso, Chile,
\ensuremath{^{bb}}Sejong University, Seoul 143-747, Korea,
\ensuremath{^{cc}}The University of Jordan, Amman 11942, Jordan,
\ensuremath{^{dd}}Universite catholique de Louvain, 1348 Louvain-La-Neuve, Belgium,
\ensuremath{^{ee}}University of Z\"{u}rich, 8006 Z\"{u}rich, Switzerland,
\ensuremath{^{ff}}Massachusetts General Hospital, Boston, MA 02114 USA,
\ensuremath{^{gg}}Harvard Medical School, Boston, MA 02114 USA,
\ensuremath{^{hh}}Hampton University, Hampton, VA 23668, USA,
\ensuremath{^{ii}}Los Alamos National Laboratory, Los Alamos, NM 87544, USA,
\ensuremath{^{jj}}Universit\`{a} degli Studi di Napoli Federico I, I-80138 Napoli, Italy
}
\noaffiliation
\date{\today}

\begin{abstract}
\noindent A measurement of vector boson ($V$) production in conjunction with a $\dstar$ meson is presented.  Using a data sample corresponding to $9.7\, {\rm fb}^{-1}$ of 
proton-antiproton collisions at center-of-mass energy $\sqrt{s}=1.96\rm~ TeV$ produced by the Fermilab Tevatron, 
we reconstruct $V+\ds$ samples with the CDF~II detector.  The \ds\  is fully reconstructed in the 
$\dstar \rightarrow D^{0}(\to K^-\pi^+)\pi^+$ decay mode.
This technique is sensitive to the associated production of 
vector boson plus charm or bottom mesons.
We measure the ratio of production cross sections $\sigma(W+D^{*})/\sigma(W)$ = $[1.75\pm 0.13 {\rm (stat)}\pm 0.09 {\rm (syst)}]\% $ and $\sigma(Z+D^{*})/\sigma(Z)$ = $[1.5\pm 0.4 {\rm (stat)}
\pm 0.2 {\rm (syst)}]\% $ and perform a differential measurement of $d\sigma(W+D^{*})/dp_T(D^{*})$.  
Event properties are utilized to determine the fraction of $V+\dstar$ events originating from different production processes. The results are in agreement with the predictions obtained with the {\sc pythia} program, limiting possible contribution from non-standard-model physics processes.

\end{abstract}




\maketitle



In high-energy proton-antiproton $(\bar{p}p)$ collisions, production of a vector boson, $V$, either a $W^\pm$ or $Z^0$ boson, can occur in conjunction with one 
or more heavy quarks, $Q$, where $Q=c, b$.  Standard model processes that contribute to $V+Q$ final states include direct (nonresonant) production as well as $V+Q$ final 
states originating from decays of
top quarks and Higgs bosons.  
In addition, high mass, 
non-standard-model states could also contribute, and 
therefore a detailed study of $V+Q$ processes could be sensitive to these contributions.    

In the standard model, production and decay of $V+Q$ states is governed by a combination of the strong and weak interactions.  For example, the $p\bar{p}\rightarrow W+c$ process is sensitive
to the magnitude of the Cabibbo-Kobayashi-Maskawa matrix element $V_{\mathit cs}$ as well as  the strange quark distribution function of the proton
\cite{StrangeQuarkPartonDist_2007,WPlusCharm_1Jet_1993}. Measurements of this process, along with the related  processes  $V+b\bar{b}$ and $V+ c \bar{c}$, can be used in conjunction with our knowledge 
of the weak interaction in order to test 
our understanding of the strong interaction at high energy 
\cite{WPlusHF_CDF1996}. 
Study of these processes 
also helps to provide  
better modeling of the backgrounds relevant to non-standard-model searches at the Tevatron and LHC.  



Previous Tevatron analyses of $p\bar{p}\rightarrow V+c$ production have measured
the absolute and relative cross sections $\sigma(V+c)/\sigma(V+\text{jets})$ \cite{WPlusCharm_CDF2008,WPlusCharm_CDF2013,D0WPlusCharm_PLB,D0ZPlusCharm_PRL}.
Restrictions on jet acceptance along with the heavy flavor identification techniques limited these analyses to high-momentum 
($p_{T}(\text{jet})\gtrsim 20\text{ GeV/c}$) charm and bottom hadrons.  In addition, systematic uncertainties arose in the   
inclusive tagging techniques used to separate light-quark, 
gluon, and heavy-quark contributions.  Similar measurements have been
carried out at the LHC \cite{cms, atlas, lhcb}.

Here, we present a complementary technique in which the
charm decay is fully reconstructed as a \dstar\  meson (denoted as \ds) through the decay chain 
$\ds \rightarrow D^{0}\pi_{s}^+$ followed by $D^0\rightarrow K^-\pi^{+}$ \cite{cc}.
The subscript on $\pi_{s}$ is used to denote a ``soft'' pion with low average momentum 
compared to the other pion arising from the $D^0$  decay.
This full charm reconstruction provides additional information (charge correlation between decay products, 
impact parameter) to further classify signal contributions.
With this identification technique, an improved measurement range of transverse momentum $p_{T}(\ds)>3\, {\rm GeV}/c$ 
is achieved, 
with an average value of  $10\, {\rm GeV}/c$.
In addition to $V+c$, the sample identified with this technique contains contributions
from $V+c\bar{c}$ and charm from the sequential decay of the $V+b\bar{b}$ process.  Full reconstruction of charmed
final states in $W$ boson events has been carried out at the LHC \cite{atlas}.

In reconstructing low-momentum charm hadrons in vector boson events, this work isolates a data sample that 
has been previously unexplored.  To date, anomalous charm hadron production in $\overline{p}p$ collisions, particularly at low $p_T(c)$, 
has not been ruled out.
By isolating and quantifying a unique set of $V+c$ events, this analysis constitutes a new test of the standard model.

\begin{figure*}
\begin{centering}
\includegraphics[width=3.25in]{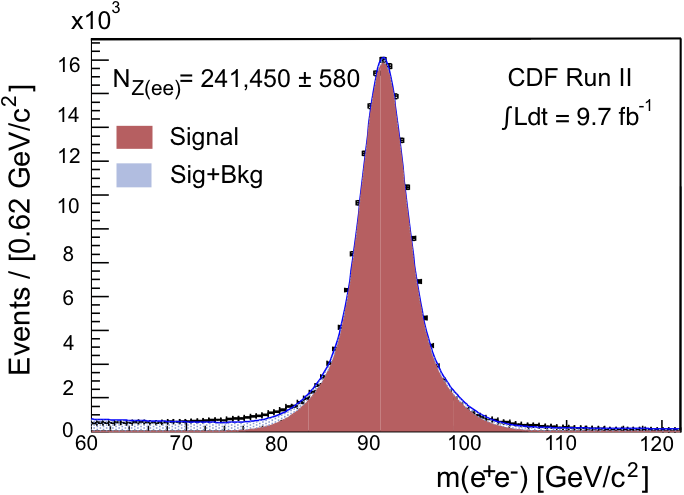}$\qquad$\includegraphics[width=3.25in]{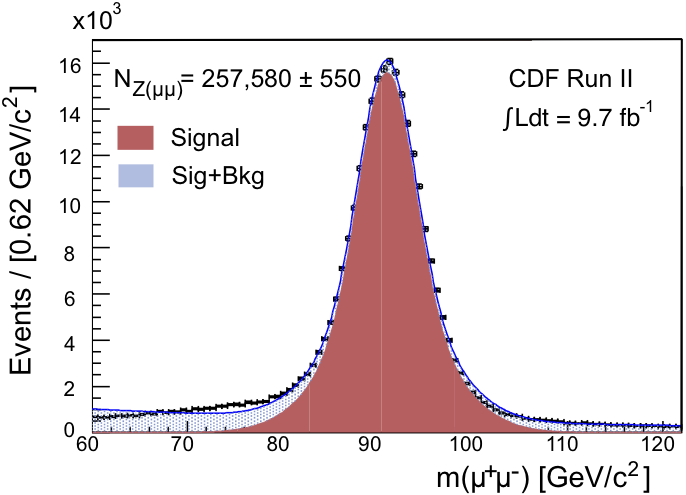}
\par\end{centering}

\caption{Distribution of dielectron (left) and dimuon (right) invariant mass. The fit to the data is described in the text.\label{fig:Zmass}}
\end{figure*}

This work utilizes the full data set collected by the CDF~II detector at the Fermilab Tevatron collider, with an integrated
luminosity of $9.7\text{ fb}^{-1}$ for $p\bar{p}$ collisions at $\sqrt{s}=1.96\text{ TeV}$.
The CDF II detector~\cite{CDFTechnicalDesignReport} is a general-purpose
magnetic spectrometer surrounded by a projective-tower calorimeter and a muon detector. The CDF II central tracking volume 
consists of a silicon detector surrounded by a large open-cell drift chamber. The entire tracking volume is contained within a uniform axial magnetic field parallel to the proton beam direction~\cite{coord}. 

Events are selected with a three-level online event selection system (trigger).  This analysis considers events selected by an inclusive high-$p_T$ lepton trigger requiring an electron (muon) with $E_{T}>18\text{ GeV}$
($p_{T}>18\text{ GeV/c}$)~\cite{coord}. 
From this high-$p_{T}$ lepton data set, we select vector-boson events as described below.

A $Z^0$ candidate is required to have two oppositely-charged, isolated 
electrons (muons), each with $E_{T}>25\text{ GeV}$ ($p_{T}>20\text{ GeV/c})$ 
and $|\eta|<1.1$.  
A lepton is considered isolated if $\iso<0.1$, where \iso \   is the isolation, defined as ratio of the
total transverse energy in a cone of radius $\Delta R\equiv\sqrt{\Delta \phi^2 + \Delta \eta^2}=0.4$
about the lepton track (excluding the lepton) 
with respect to $E_T$($p_T$) of the electron (muon).
The invariant mass of the lepton pair 
$m(\ell^+ \ell^-)$ 
is
required to fall between $66$ and $116\, {\rm GeV}/c^2$. 
Background in the $Z^0$ candidate sample is estimated by fitting the invariant
mass 
distribution to a double-Gaussian signal
plus an exponential background hypothesis. The number of signal events is obtained by integrating the double-Gaussian distribution, yielding $241\, 450 \pm 580$ ($257\ 580 \pm 550$) $Z^0\rightarrow ee (\mu\mu)$ events, where the uncertainty is statistical only.  The dilepton data and fits are shown in Fig.~\ref{fig:Zmass}.

A candidate $W^+$ event is required to have missing transverse energy $\met >25(20)\text{ GeV}$ \cite{met},
one isolated electron (muon) with $E_{T}>25\text{ GeV}$ ($p_{T}>20\text{ GeV}/c)$ and pseudorapidity $|\eta|<1.1$. 
Missing transverse energy   is corrected for
muons and other instrumental effects 
that may produce false \met.
Finally, the transverse mass of the $W$, $M_{T}=\sqrt{2 E_{T} \met (1-\cos\Delta\phi})$, where 
$E_T$ is the transverse energy of the lepton and 
$\Delta \phi$ is the azimuthal angle between the lepton and missing transverse energy,
must satisfy $M_{T}>20\text{ GeV}/c^2$.

\begin{figure*}
\begin{centering}
\includegraphics[width=3.25in]{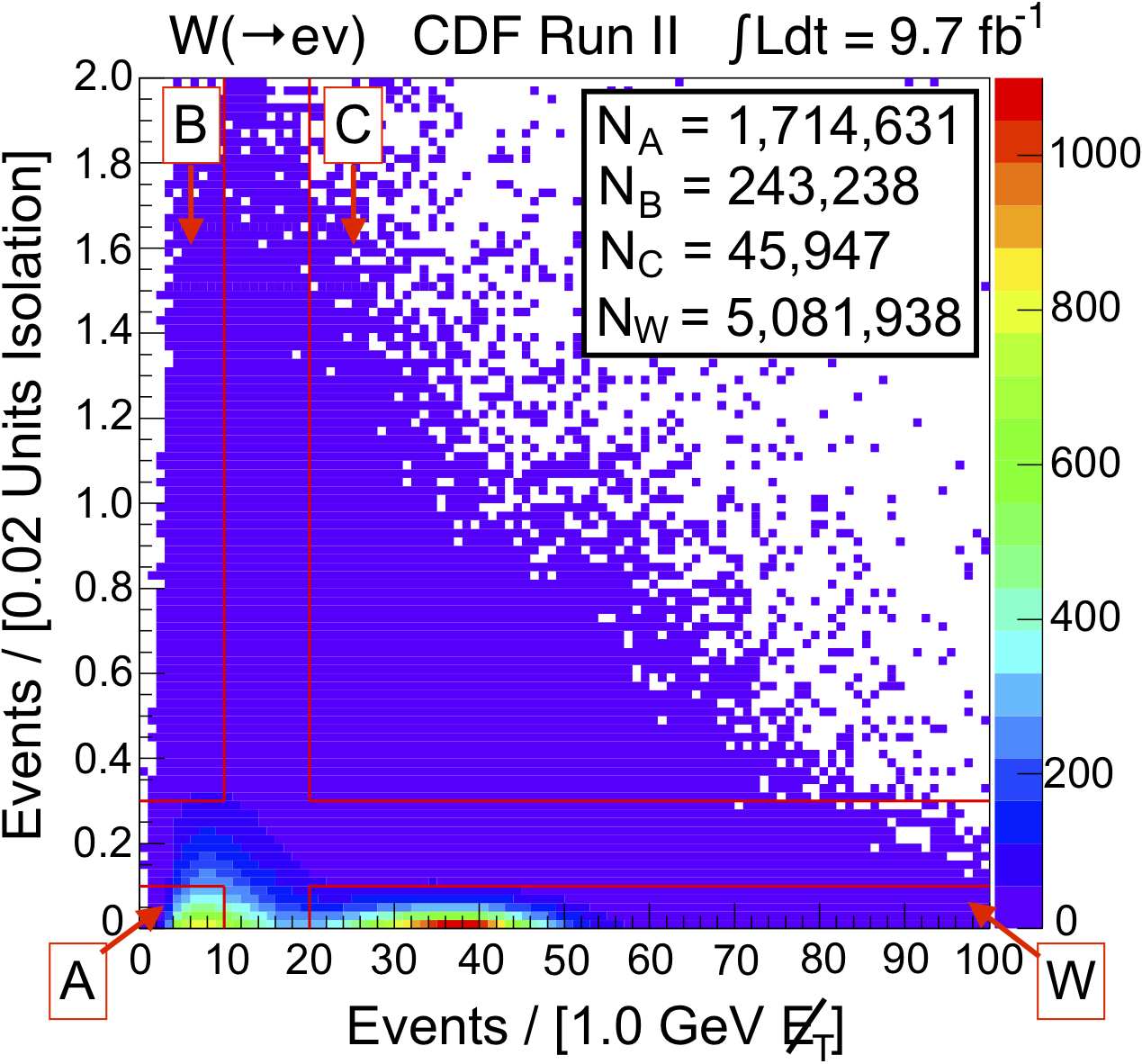}$\qquad$\includegraphics[width=3.25in]{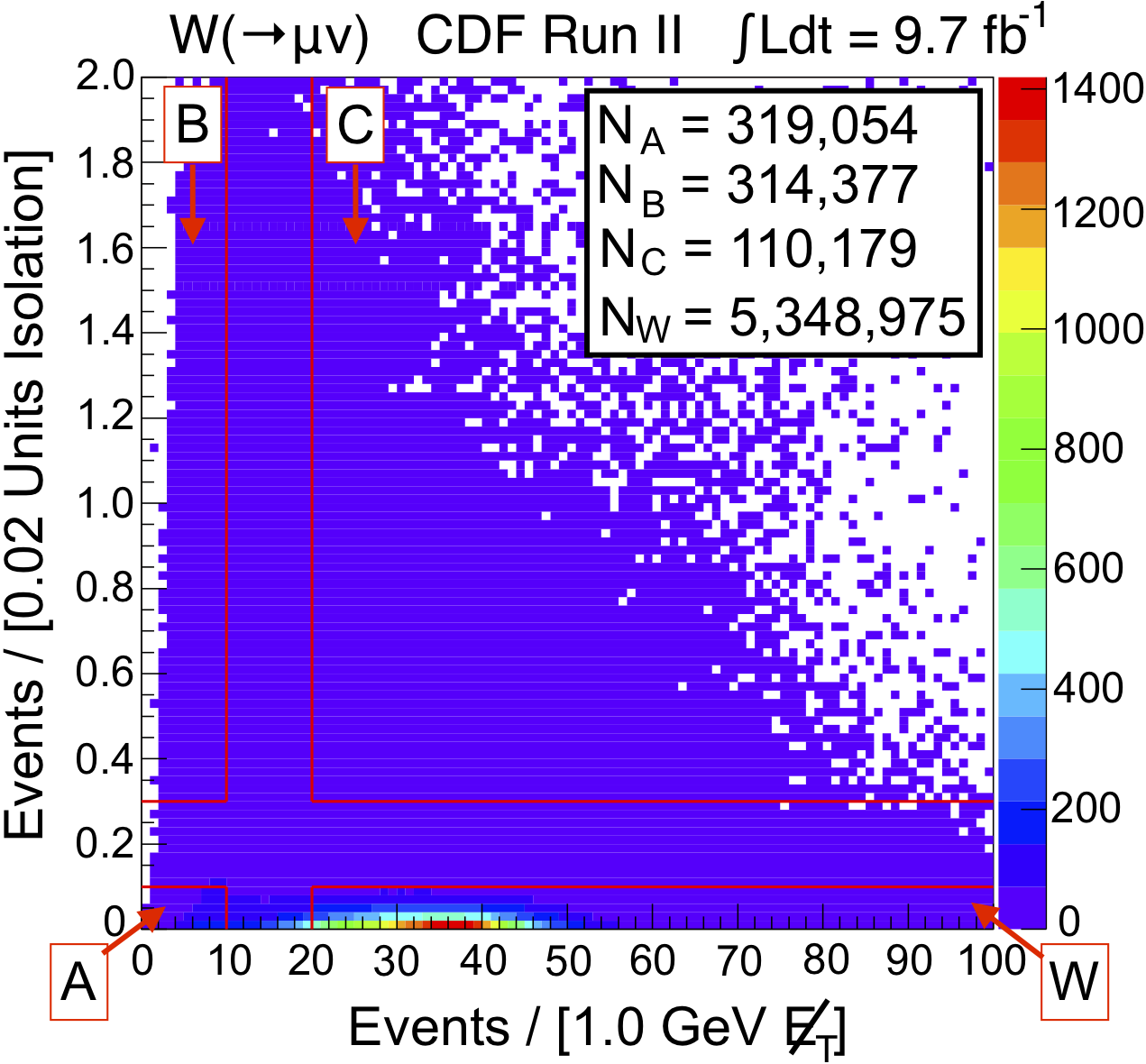}
\par\end{centering}

\caption{Lepton isolation versus missing transverse energy for electron (left) and muon (right) events passing the selection criteria described in the text.  The regions 
identified in Table~\ref{tab:BoundaryDefs} are shown graphically on the plots.  The $W$ signal is dominant in the region $I<0.05$ and $20<\met<50\, {\rm GeV}/c$.  \label{fig:metiso}}
\end{figure*}

Background to the $W$ candidates consists of electroweak sources
($Z\rightarrow\ell^+\ell^-$ and $W\rightarrow\tau\nu$ events, denoted as EWK hereafter) and from
hadrons displaying a similar topology as  $W$ events (multijet background).
We estimate the number of all such background events by first relaxing the selection criteria on
lepton \iso  \  and event \met, and then defining
four regions A, B, C, and W in the \met/\iso \   plane (Table~\ref{tab:BoundaryDefs}).
Region W is the signal region as defined earlier, while the other regions contain
primarily multijet events mimicking the $W$ signature. Lepton isolation versus missing
transverse
energy is shown in Fig.~\ref{fig:metiso}.

\begin{table}
\begin{centering}
\caption{Boundaries used to define regions A, B, C, and W 
for $W$ candidates in the \met -\iso \   plane. Regions
A, B, and C consist primarily of multijet background,
and are used to estimate the multijet background content of signal region W.
Numbers in parentheses are for $W^+ \to\mu^+ \nu$ when different from the $W^+\to e^+\nu $ boundaries. \label{tab:BoundaryDefs}}
\begin{tabular}{ccc}
\hline \hline  
Region& \met \  range & \iso \  range \\ 
\hline 
A & $<10$ GeV & $<0.1$\tabularnewline
B & $<10$ GeV &  $>0.3$\tabularnewline
C & $> 25 (20)$ GeV & $>0.3$\tabularnewline
W & $> 25 (20)$ GeV & $< 0.1$\tabularnewline
\hline \hline
\end{tabular}
\par\end{centering}

\end{table}

We estimate $N_{\rm jet}^{W}=N_{\rm jet}^{C} \times N_{\rm jet}^{A} / N_{\rm jet}^{B}$,
where $N_{\rm jet}^{X}$ is the number of multijet events in region $X$, as done in Ref.~\cite{WAndZCrossSections2007}.
The contamination arising from EWK and residual signal events in
regions A, B, and C is estimated using the {\sc pythia 6.2}  Monte Carlo simulation \cite{Pythia6p4Manual}.  
The simulation is used to predict the backgrounds
relative to the
yield of $W\rightarrow\ell\nu$ events
in signal region W. By relating other contributions to the number of signal events, 
we produce a system of linear equations to be solved for
the number of $W\rightarrow\ell\nu$ signal events in region
W.  The solutions determine the contribution of each 
background in each region. Of $5\, 081\, 938 \ (5\, 348\, 975)$  
candidate events in region W, 93.6 (91.5)\% are $W\rightarrow e\nu(\mu\nu)$ signal.
Uncertainties in these estimates are considered later in the determination
of fractions $N(W+\ds)/N(W)$.


For all $W$ and $Z$ candidate events, we search for \ds\  mesons by considering
all reconstructed charged particles (tracks) within $2.0$ cm longitudinally of the point of
closest approach of the high-$p_T$ lepton to the beamline ($|\Delta z|<2.0\text{ cm}$).
For each possible set of three tracks with unit total net charge, we hypothesize a match to the $\ds \rightarrow D^{0}(\rightarrow K^-\pi^+)\pi_{s}^+$ decay mode and test for consistency.
To ensure well-measured tracks, each track must satisfy $|\eta|<1.1$, and $p_{T}(K)$ and $p_T(\pi)>400\text{ MeV/{\it c}}$,
$p_{T}(\pi_{s})>80\text{ MeV/{\it c}}$.  We also require $\Delta R<1.1$ for each pair
of tracks. 
As $D^{0}\rightarrow K^{-}\pi^{+}$ is CKM-favored versus
$D^{0}\rightarrow K^{+}\pi^{-}$ by a factor of $10^{4}$, the two pions are  required
to have the same charge. 

The resulting sample is subjected to a kinematic fit to
reconstruct \ds\  and $D^{0}$ vertices from the $K,\pi$ and $\pi_{s}$
track candidates.  The $K$ and $\pi$ candidates must intersect to form a candidate 
$D^0$ vertex,
and the $D^0$ candidate must intersect with the $\pi_s$ to form
a $\ds$ vertex.
Since direct charm (\it e.g., \rm $p\bar{p}\to Wc$, $p\bar{p}\to Wc\bar{c}$) 
as well as indirect charm
(\it e.g., \rm $Wb\bar{b}$ with $b\to c$) are of interest, 
the trajectory of the reconstructed \ds\ is not required 
to point back to the primary $\bar{p}p$ vertex.
Finally, we require the $D^{0}$ mass, determined by the fit, to fall within $3\sigma=0.033\text{ GeV/$c^2$}$
of the known $D^{0}$ mass value, $m_{D^{0}}=1.865\text{ GeV/$c^2$}$ \cite{PDGBook2013}.



To reduce background in the \ds\  selection, we train an artificial neural
network (ANN) to discriminate among \ds\  candidates using 19 
variables as described in chapter 5 of Ref.~\cite{KMateraPhd2014}.  These variables include charged-particle momenta
and opening angles; the distance between the $p\bar{p}$ collision vertex and the reconstructed $\ds$ and $D^0$ vertices; 
and track impact parameters.
To train the ANN, we provide samples of signal and background events. For signal, we generate
a sample of $W+\ds$ events using {\sc pythia 6.2} \cite{Pythia6p4Manual} followed by a full detector simulation. 
For ANN training,  we define the signal sample as all simulated $W+\ds$ candidates with mass difference, $\Delta m\equiv m(K\pi\pi_{s})-m(K\pi)$, within $3\sigma=0.0029 \, {\rm GeV}/c^2$ of 
the known mass-difference value $\Delta m (\ds -D^0)=0.1455\, {\rm GeV}/c^2$ \cite{PDGBook2013}.
To model background for ANN training, we choose \ds\  candidates in data that pass
all fitting and kinematic selection criteria, but whose final-state particles 
do not have the proper sign correlations ({\it e.g.,} $K^-\pi^-\pi^{+}_s$). 
To model background events as accurately as possible, we require these
background \ds\   candidates to fall within $3.2\sigma$ of the
peak, a compromise that allows an equal number of signal and background
events for optimal ANN training. Control data samples as well as simulated samples are utilized to verify that the ANN is well behaved and unbiased~\cite{KMateraPhd2014}.

We map ANN output scores to the region $[-1.0,1.0]$, with more
positive numbers being more signal-like.  Candidates are required   
to have an ANN score greater than $0.0$ to be classified as signal.
This suppresses $80\%$ of the background while retaining $90\%$ of the signal.

With this algorithm, we search the selected vector-boson-candidate events for $\ds$ candidates. To measure the $\ds$ yield, 
we fit the $\Delta m$ distribution
to a power-law background plus double-Gaussian signal hypothesis. We use a template
for the double-Gaussian signal, taken from a fit to 
simulated signal events, with the signal width increased by a factor of $1.1$. This accounts for the slightly better mass resolution observed in the simulation \cite{KMateraPhd2014}.

\begin{figure*}
\begin{centering}
\includegraphics[width=3.25in]{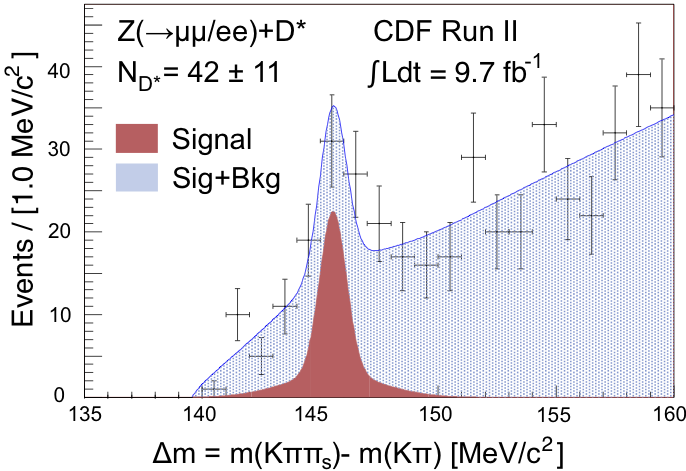}$\qquad$\includegraphics[width=3.25in]{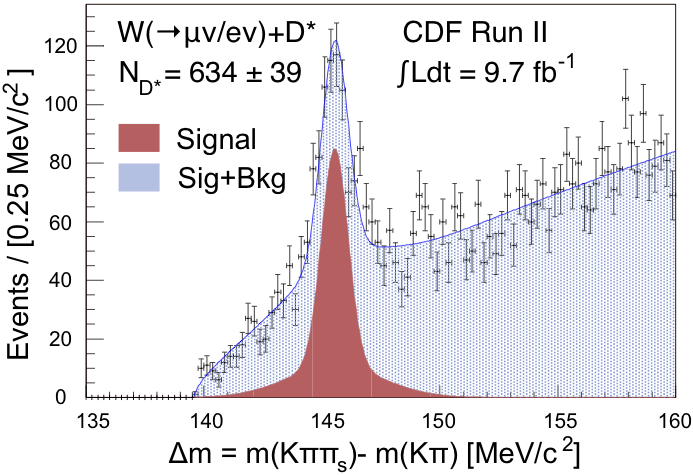}
\par\end{centering}

\caption{Distribution of mass difference between the \ds\ and $D^0$ candidates in $Z$ and $W$ events, with fit results overlaid. The electron and muon decay
modes are combined. \label{fig:WZDeltaMFittedPlots_postNN}}
\end{figure*}

We obtain the following yields in the $W$ candidate sample: 
$N(W(e\nu)+\ds)=340\pm30$ and $N(W(\mu\nu)+\ds)=294\pm26$, where the uncertainties are statistical only.  
In the $Z$ candidate sample, we observe 
$N(Z(ee)+\ds)=22\pm9$ and $N(Z(\mu\mu)+\ds)=20\pm7$ events.
When the electron and muon decay channels are combined, the yields are $N(W(\mu\nu/e\nu)+\ds)=634\pm39$ and $N(Z(\mu\mu/ee)+\ds)=42\pm11$. 
The mass-difference distributions are shown in Fig. \ref{fig:WZDeltaMFittedPlots_postNN}.
Yields reported here are the number of reconstructed \ds \  in events passing $V$ candidate selection criteria, and therefore represent
the number of reconstructed \ds \ produced in conjunction with both signal and background $V$ candidates.
In the following, we account for the background $V$ contribution in $V+\ds$ candidate events.

For $Z$ candidates, we define two regions in $\ell^{+}\ell^{-}$ mass: The signal region is defined as 
$|m_{\ell^+\ell^-}-91\ {\rm GeV}/c^2|\le3\sigma$,
and the background region is defined as $|m_{\ell^+\ell^-}-91\ {\rm GeV}/c^2|>3\sigma$, with
$\sigma=2.0\text{ }(3.0)\ {\rm GeV}/c^{2}$ for $Z\rightarrow\mu\mu( ee)$ bounded by the $66 < m_{\ell^+\ell^-}<116\ {\rm GeV}/c^2$ mass window.
We fit the $m_{Z}$ distribution to a double-Gaussian signal plus
exponential background hypothesis, and integrate beneath the curves
to estimate the signal and background yields in each region.
For each region, we fit the 
$\Delta m$ for all \ds\  candidates using our previously defined signal plus background fitting function.
This provides \ds\ yields in the signal and background regions.
Two coupled linear equations are used to solve for the
rates $f_{\ds}^{\mathrm Z,sig}$ ($f_{\ds}^{\mathrm Z,bkg}$) at which signal (background) $Z$ candidates are produced with \ds\  mesons as
described in chapter 6 of Ref.~\cite{KMateraPhd2014}.

After measuring these rates, the procedure is repeated for various choices of signal and background
$m(\ell^+\ell^-)$ boundaries. We fit the set of all $f_{\ds}^{\mathrm Z,sig}$ values found using
these definitions to a constant value hypothesis, and take the variation in this fit to be the systematic uncertainty due to the unknown rate of \ds \ 
mesons produced in association with $V$ backgrounds. This is done separately
for $Z\to e^+e^-$ and $\mu^+\mu^-$ decay channels, giving a systematic uncertainty in $f_{\ds}^{\mathrm Z,sig}$ of $20\%$ for the
$Z\rightarrow\mu\mu$ channel, and of $11\%$ for the $Z\rightarrow ee$
channel. 


To determine the number of real $V$ plus real \ds \ events, we follow a procedure similar to that outlined above to 
determine the background contributions to the $W$ sample.  We split candidate events 
into \met/\iso \  regions A, B, C, and W, as defined in Table~\ref{tab:BoundaryDefs}. Each region contains \ds\ mesons 
produced with signal $W$, as well as \ds\ produced with three background sources: $W\rightarrow\tau\nu$, $Z\rightarrow\ell^+\ell^-$, and multijet events. 
The linear equations are solved simultaneously for the rates $f_{\ds}^{\mathrm W,sig}$ ($f_{\ds}^{\mathrm W,bkg}$) at 
which signal (background) $W$ candidates are produced with \ds \ mesons (see Section~6.3 of Ref.~\cite{KMateraPhd2014}.) The associated 
systematic uncertainty is determined by repeating the procedure for several definitions of regions A, B, C and W, first keeping the
\iso \  boundaries fixed and varying the \met \  boundaries,
and then keeping the \met \  boundaries fixed and varying
the \iso \  boundaries. This is done separately for the electron and muon $W$
decay channels. The resulting 
systematic uncertainty in $f_{\ds}^{W}$ is $10\%$ in the $W\rightarrow e\nu$ channel, 
and $2\%$ in the $W\rightarrow\mu\nu$ channel. The larger uncertainty for the 
$W(\rightarrow e\nu)$ case is due to a larger
fraction of multijet background, relative to the $W\rightarrow \mu \nu$ case.

The full results of the sample composition studies are reported in chapter 6 of Ref.~\cite{KMateraPhd2014}. We observe the  
rate of \ds \ mesons produced in events with falsely reconstructed  $V$ bosons to be considerably higher than the rate of \ds \ mesons observed in real $V$ events.  This arises through Quantum Chromodynamic (QCD) processes producing falsely 
reconstructed leptons. 


We unfold tagged signal fractions, $f_{\ds}^{V}$, to ratios $\sigma(V+\ds)/\sigma(V)$, where $V=(W,Z)$, by following the prescription
\begin{equation}
\frac{\sigma\big(V+\ds\big)}{\sigma\big(V\big)}=\frac{f_{\ds}^{V}(A\cdot\epsilon)_{V}}{(A\cdot\epsilon)_{V+\ds} {\cal B}(\ds\rightarrow D^{0}(\rightarrow K\pi)\pi_{s})}\label{eq:Unfolding_PublicNote}
\end{equation}
Here, the branching ratio ${\cal B}(\ds\rightarrow D^{0}(\rightarrow K\pi)\pi_{s})$  is $0.0263\pm0.0004$
\cite{PDGBook2013}, and $(A\cdot\epsilon)_{V}$, $(A\cdot\epsilon)_{V+\ds}$  is our total (geometric plus kinematic)
 acceptance for $V$ and $V+\ds$ events, respectively.

\begin{figure}
\begin{centering}
\includegraphics[width=3.25in]{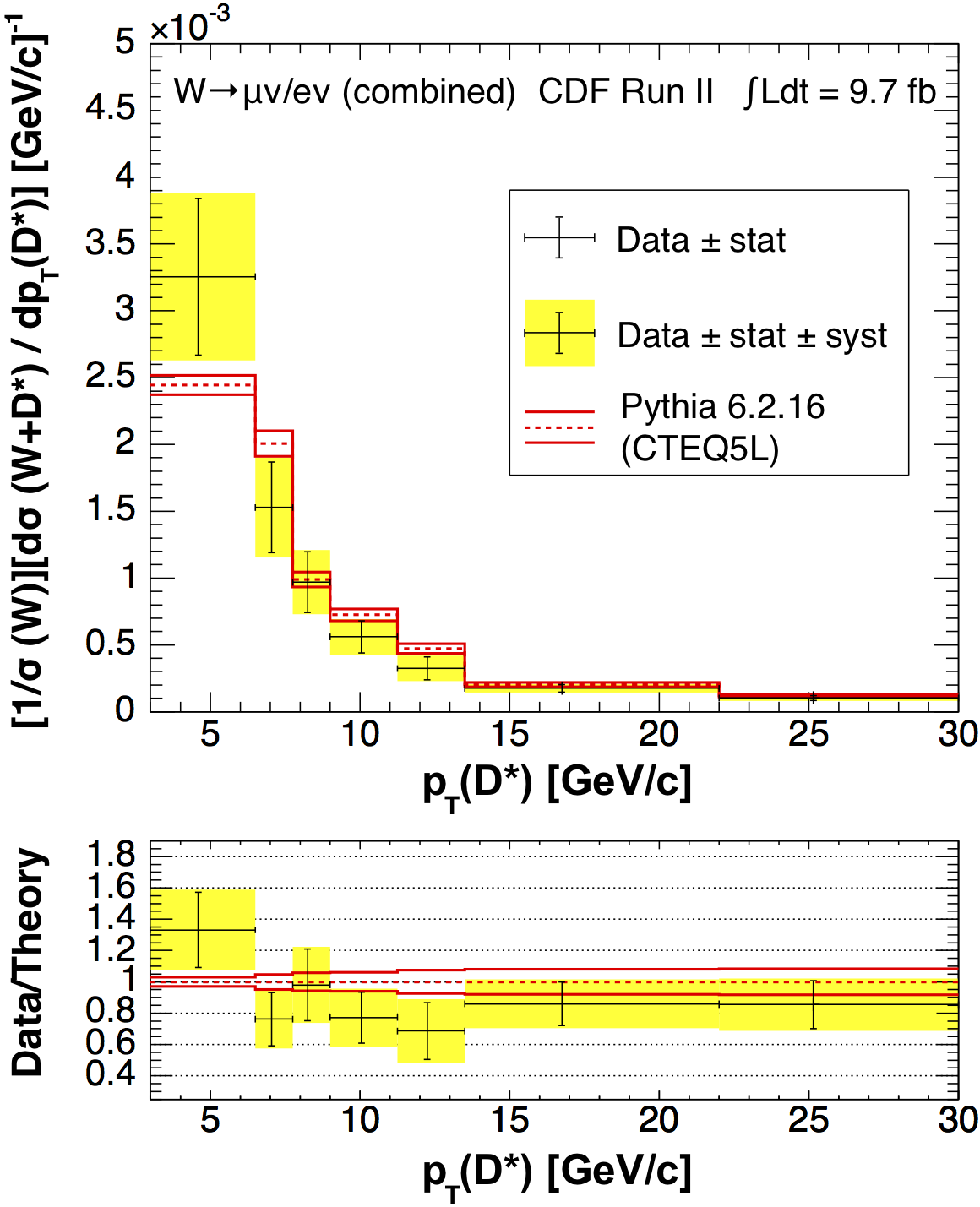}
\par\end{centering}

\caption{Ratio of cross sections $\sigma(W+\ds)/\sigma(W)$
as a function of $p_{T}(\ds)$, for combined $W\rightarrow e\nu$
and $W\rightarrow\mu\nu$ results. Error bars show the statistical uncertainty; the sum in quadrature
of the statistical and systematic errors is shown as a yellow error
band. The dotted red line shows the prediction of {\sc pythia 6.2} obtained
using the CTEQ5L PDF, with solid red lines showing PDF uncertainty
in this prediction. The ratio of the simulated distribution to data
is shown in the lower plot.\label{fig:Pretty_Unfolded_Fractions_Diff_W_Combined}}
\end{figure}

To determine the $V$ acceptances, we apply the $W/Z$ tagging algorithms
over inclusive simulated samples generated using {\sc pythia 6.2}. For the $V+\ds$ acceptances, we determine 
the values of ($A\cdot\epsilon$)
both differentially as a function of $p_T(\ds)$, and for the inclusive set of all events with $p_{T}(\ds)>3\text{ GeV}$, the threshold below which the analysis acceptance vanishes. 
In the differential case, bins are chosen such that the expected number of tagged \ds\ candidates  
in each bin is approximately constant according to simulation. Bins are as shown in Fig.~\ref{fig:Pretty_Unfolded_Fractions_Diff_W_Combined};
refer to chapter 8 of Ref.~\cite{KMateraPhd2014} for specific values.


Systematic uncertainties on the production cross sections arise from sample yield and purity 
estimates, acceptance, efficiency, and simulation~\cite{KMateraPhd2014}.  
Sources of uncertainty that are common in measurements of total cross section
(\emph{e.g.}, luminosity and trigger efficiency uncertainties), cancel
when taking the ratio $\sigma(V+\ds)/\sigma(V)$.

The systematic uncertainty in ($A\cdot\epsilon$) is dominated
by
the uncertainty in the CTEQ \cite{cteq} parton-distribution function (PDF) of the proton chosen for this analysis.
Uncertainties due to the PDFs are estimated using 90\% confidence-level (C.L.) variations 
on the 20 CTEQ eigenvectors. To investigate possible further dependencies on the chosen PDF, the MSTW2008~\cite{mst2008} central value
is also checked.  When the difference between
CTEQ and MSTW2008 is smaller than the uncertainty from variations on the 20 eigenvectors,
no additional uncertainty is taken. When the difference is larger, that
difference is added in quadrature
to the eigenvector uncertainty. Uncertainty in $\alpha_{s}(M_{Z})$
at $90\%$ C.L. is taken as an additional systematic uncertainty, and added in
quadrature to the above.

For both the $W$ and $Z$ samples, we find the ratio $\sigma(V+\ds)/\sigma(V)$ for the inclusive set of events
$p_{T}(\ds)>3\text{ GeV}/c$.  For the $W$ samples only, we unfold
$f_{\ds}^{W}$ differentially as a function of $p_{T}(\ds)$.
Results from the electron and muon decay channels are combined
using a best-linear-uncertainty estimator, assuming that systematic uncertainties are
fully correlated across $W$ or $Z$ decay modes. 
The final results are shown in Table~\ref{tab:FinalRatios} and 
Fig.~\ref{fig:Pretty_Unfolded_Fractions_Diff_W_Combined}, 
and in all cases 
agree with the {\sc pythia} predictions within uncertainties.

\begin{table}
\begin{centering}
\caption{Ratio of cross sections $\sigma(V+\ds)/\sigma(V)$ for the inclusive
sample $p_{T}(\ds)>3\ \text{ GeV}/c$, and predictions of {\sc pythia 6.2.16} 
simulation using the CTEQ5L PDF.  For results from data, the first uncertainty is statistical,
the second is systematic.  For results from simulation, uncertainty listed is statistical.  
\label{tab:FinalRatios}}
\begin{tabular}{lcc}
\hline \hline
& \multicolumn{2}{c}{$\sigma(V+\ds)/\sigma(V)$ (in \%)} \\
Sample & Data  & {\sc pythia 6.2.16} \tabularnewline
\hline 
$W\rightarrow e\nu$ & $1.74\pm0.21\pm0.17$ & \tabularnewline
$W\rightarrow\mu\nu$ & $1.75\pm0.17\pm0.05$ & $1.77\pm0.07$\tabularnewline
$W$ combined & $1.75\pm0.13\pm0.09$ & \tabularnewline
$Z^0\rightarrow ee$ & $1.0\pm0.6\pm0.2$ & \tabularnewline
$Z^0\rightarrow\mu\mu$ & $1.8\pm0.5\pm0.2$ & $1.36\pm0.05$\tabularnewline
$Z^0$ combined \ \ \   & $1.5\pm0.4\pm0.2$ & \tabularnewline
\hline \hline
\end{tabular}
\par\end{centering}

\end{table}

The $W+\ds$ signal events are expected to
come from three different production processes: $s(d)+g \rightarrow W+c$ (\emph{Wc}), 
$q+\bar{q}^{\prime} \rightarrow W+g(\rightarrow c\bar{c})$ (\emph{Wcc}), and $q+\bar{q}^{\prime} \rightarrow W+g(\rightarrow b\bar{b})$ (\emph{Wbb}). Due primarily to a difference in $p_{T}(\ds)$ spectra, a neural
network may generally identify \ds\  from different sources (\emph{Wcc},
\emph{Wc}, and \emph{Wbb}) with different efficiencies. This enables the fraction
of signal from each production process to be determined.

We train two tiers of neural networks; the first consists of networks that
are trained to identify one type of signal (\emph{Wcc}, \emph{Wc, }and \emph{Wbb})
versus the $\ds$ background sample described earlier; the second consists
of networks that are trained to identify one type of signal over another
(\emph{Wcc} vs \emph{Wbb}, \emph{Wc} vs \emph{Wcc}, and \emph{Wbb} vs \emph{Wc}).
Each first-tier neural network is paired with a second-tier neural network, in
order to select $\ds$ candidates from one production source preferentially, 
over all other $\ds$ candidates as described in chapter 10 of Ref.~\cite{KMateraPhd2014}.
We apply each ANN-pair to simulated signal events
to determine the efficiencies with which each pair identifies $\ds$ from each production source,
obtaining a matrix of efficiencies. We finally apply  each ANN-pair to the data and determine three yields.
We solve the resulting linear equations to find fractions $X^{Wc}$, $X^{Wcc}$, and $X^{Wbb}$, where
$X^{Y}$ is the fraction of $W+\ds$ signal that comes from production process $Y$.  After
repeating the technique by varying the efficiencies for each process, we determine that
the uncertaintly is dominantly statistical.

This method gives good separation for $X^{Wbb}$, but is not very
effective at determining the fraction $X^{Wc}$ owing to similarities
in the kinematic properties of \ds\  from $Wcc$, and from $Wc$.  
In the case of \emph{Wc} production,
conservation of charge requires that the $W$ and $\ds$ are produced
with opposite signs. In \emph{Wcc} or \emph{Wbb} production, however,
we are equally likely to identify oppositely charged $W$ and \ds, $(W+\ds)_{\mathrm{OS}}$,
or same-signed $W$ and \ds, $(W+\ds)_{\mathrm{SS}}$.
We therefore determine the number of \emph{Wc} events in the $W+\ds$ signal by taking the difference 
$(W+\ds)_{\mathrm{OS}}-(W+\ds)_{\mathrm{SS}}$. We divide by the sum $(W+\ds)_{\mathrm{OS}}+(W+\ds)_{\mathrm{SS}}$ 
to get the fraction $X^{Wc}$.  The uncertainty in this measurement of $X^{Wc}$ is statistical only and
treated as uncorrelated with the uncertainty
in the measurement of $X^{Wbb}$, due to the use of a different technique for finding
each fraction. Normalized to unity, the production-process fractions in the 
overall $W+\ds$ signal are obtained as 
$X^{Wcc}=73\pm8\%$, $X^{Wbb}=13\pm5\%$, $X^{Wc}=14\pm6\%$.  The relative contribution of these processes is momentum 
dependent.  In particular, our ability to identify low-momentum \ds \ candidates enhances the $Wcc$ contribution 
compared to analyses 
performed at higher momenta \cite{KMateraPhd2014}.


In conclusion, we present the first measurement of 
$V+\ds$ production at hadron colliders in the regime $p_{T}(\ds)>3\text{ GeV}/c$. 
The expected rate of \ds\  production in $V$ events is as predicted
by {\sc pythia 6.2}, both for the integrated sample $p_{T}(\ds)>3\text{ GeV}/c$,
and differentially as a function of $p_{T}(\ds)$.

We thank the Fermilab staff and the technical staffs of the
participating institutions for their vital contributions. This work
was supported by the U.S. Department of Energy and National Science
Foundation; the Italian Istituto Nazionale di Fisica Nucleare; the
Ministry of Education, Culture, Sports, Science and Technology of
Japan; the Natural Sciences and Engineering Research Council of
Canada; the National Science Council of the Republic of China; the
Swiss National Science Foundation; the A.P. Sloan Foundation; the
Bundesministerium f\"ur Bildung und Forschung, Germany; the Korean
World Class University Program, the National Research Foundation of
Korea; the Science and Technology Facilities Council and the Royal
Society, United Kingdom; the Russian Foundation for Basic Research;
the Ministerio de Ciencia e Innovaci\'{o}n, and Programa
Consolider-Ingenio 2010, Spain; the Slovak R\&D Agency; the Academy
of Finland; the Australian Research Council (ARC); and the EU community
Marie Curie Fellowship Contract No. 302103.


\begin{thebibliography}{10}
\bibitem{cc} Charge conjugate modes are implied throughout unless specifically noted.
\bibitem{coord} A coordinate system is used with the $z$ axis along the proton beam direction; $\theta$ is the polar angle, $\phi$ is the 
azimuthal angle, pseudorapidity is $\eta = - \ln[\tan({\theta/2})]$, tranverse momentum is $p_T=|p|\sin\theta$, and transverse energy
is $E_T = E \sin\theta$.
\bibitem{met} The missing tranvserse energy, \met , is defined as the magnitude of the energy imbalance as measured in the calorimeter, 
$E\!\!\!/_T = |-\sum{\vec{E}_T}|$.  The \met is corrected for calorimeter geometry, highly penetrating muons and other instrumental effects.
\end{thebibliography}

\end{document}